\documentclass[11pt]{article}
\usepackage{amsmath, graphics,amssymb,epsfig,subfigure}
\usepackage{algorithm}
\usepackage{algorithmic}
\usepackage{float}
\usepackage{multirow}
\usepackage{cuted,flushend}
\usepackage{midfloat}

\newcommand{\qed}{\nobreak \ifvmode \relax \else
      \ifdim\lastskip<1.5em \hskip-\lastskip
      \hskip1.5em plus0em minus0.5em \fi \nobreak
      \vrule height0.75em width0.5em depth0.25em\fi}
\begin{document}
\title{Aligned Interference Neutralization\\ and the Degrees of Freedom of the $2$ User Interference Channel with Instantaneous Relay}
\author{\normalsize  Namyoon Lee$^\dagger$, Syed A. Jafar$^\ddagger$\bigskip
\\
$^\dagger$ \small Samsung Electronics CO., LTD, Suwon, Korea\\
$^\ddagger$ \small University of California Irvine\\
      {\small \it E-mail~:~namyoon.lee@gmail.com, syed@uci.edu} \\
       }
\date{}

\maketitle
\begin{abstract}
It is well known that the classical 2 user Gaussian interference channel has only 1 degree of freedom (DoF), which can be achieved by orthogonal time division among the 2 users. It is also known that the use of conventional relays, which introduce a processing delay of at least one symbol duration relative to the direct paths between sources and destinations, does not increase the DoF of the 2 user interference channel. The use of instantaneous relays (relays-without-delay) has been explored for the single user point-to-point setting and it is known that such a relay, even with memoryless forwarding at the relay, can achieve a higher capacity than conventional relays. In this work, we show that the 2 user interference channel with an instantaneous relay, achieves $\frac{3}{2}$ DoF. Thus, an instantaneous relay increases not only the capacity but also the DoF of the 2 user interference channel. The achievable scheme is inspired by the aligned interference neutralization scheme recently proposed for the $2\times 2\times 2$ interference channel. Remarkably the DoF gain is achieved with memoryless relays, i.e., with relays that have no memory of past received symbols.


\end{abstract}
\newpage

\section{Introduction}
Due to the broadcast nature of the wireless medium, simultaneously
transmitted signals from multiple source nodes cause interference
to each other. The interference gives rise to
competition between different information flows, but it can also be
exploited for cooperation between them. Understanding the complex interplay between cooperation and competition is the key to understanding the capacity limits of wireless networks. In spite of an abundance of literature studying cooperative interference networks, the understanding of this topic --- even at a coarse degrees-of-freedom (DoF) level ---   is far from mature, as evident from any sampling of recent works on DoF characterizations of cooperative interference networks that have produced surprising insights. Continuing in the same spirit, in this work we seek to illuminate a few  interesting and fundamental aspects of the interplay between cooperation and competition, from a DoF perspective. The setting we wish to explore is the 2 user interference channel with an instantaneous relay.

\subsubsection*{\it Competition for DoF in the 2 User Interference Channel }
From a DoF perspective, the two user interference channel represents competition for resources in the strongest sense ---  whatever DoF one user is able to achieve, comes at the cost of an equal amount of the other user's DoF. This is because it is well known that the 2 user interference channel has only 1 DoF, which can be achieved by any orthogonal time division policy between the two users.

\subsection*{\it The Use of Relays in the Interference Channel}
To alleviate the competition between the two users, the use of relay nodes has been considered. Indeed, there are several results that show increased rates in the form of SNR gains from employing a relay in an interference channel \cite{Sahin1, Sahin2, Maric1, Maric2, Tian}. DoF gains are achieved with the use of relays when source and destination nodes are not directly connected, e.g., in \emph{layered} relay networks. Reference \cite{Ilan_Avestimehr} provides a characterization of the DoF of the multihop 2 user interference channel with layered relays. However, if direct channel coefficients between source and destination nodes are non-zero, then from a DoF perspective, it has been shown that the use of relays does \emph{not} contribute any DoF gain for the 2 user interference channel \cite{Cadambe2}, regardless of the number of relay nodes. Remarkably, this is true even if the relay nodes are equipped with multiple antennas\footnote{Exceptions include cases where relays are cognitive, i.e., they have prior knowledge of source messages \cite{Sridharan}, or when the power constraints at the relay nodes are  orders of magnitude higher than the source nodes \cite{Tannious}.}.

\subsection*{\it From Conventional Relay to Instantaneous Relay}
Since conventional relays are shown not to increase the DoF of a two user interference channel, we explore another form of relaying, introduced by El Gamal and Hassanpour \cite{Gamaal}, known as \emph{instantaneous relaying} (\emph{relay-without-delay}). Unlike the classical relay whose transmitted signal is a function of only past observations, the transmitted signal from an instantaneous relay, is allowed to depend on both past and \emph{current} received signals. Furthermore, a \emph{memoryless} instantaneous relay is a constrained form of an instantaneous relay, where the transmitted signal is allowed to depend \emph{only on the current} received signal. Clearly, a memoryless instantaneous relay is a simpler but less powerful relay than a general instantaneous relay, which in turn is more powerful than a conventional relay. Interestingly, no order relationship necessarily exists between conventional relays and instantaneous memoryless relay. Because the latter has the benefit of being able to react to the current received signal but the handicap of having no memory of past received signals, it may or may not be better than a conventional relay in general. For the single user setting,  since the memoryless nature of the relay does not allow decode and forward processing, it is clear that such a relay channel will have a smaller capacity than a conventional relay in those settings where decode and forward processing is needed, e.g., when there is no direct path between the source and destination. On the other hand, El Gamal and Hassanpour \cite{Gamaal} show that the memoryless instantaneous relay channel can have a higher capacity than a conventional relay. While the result is surprising, the capacity gains are modest and limited to SNR gains. However, as we show in this paper, the use of a memoryless instantaneous relay in a 2 user interference channel has a very significant impact as it increases the DoF by at least 50\% (from 1 to 1.5) for almost all channel realizations.

\subsection*{\it Practical Motivation for Interference Channel with Instantaneous Relay}
\begin{figure}
\centering
\includegraphics[width=4in]{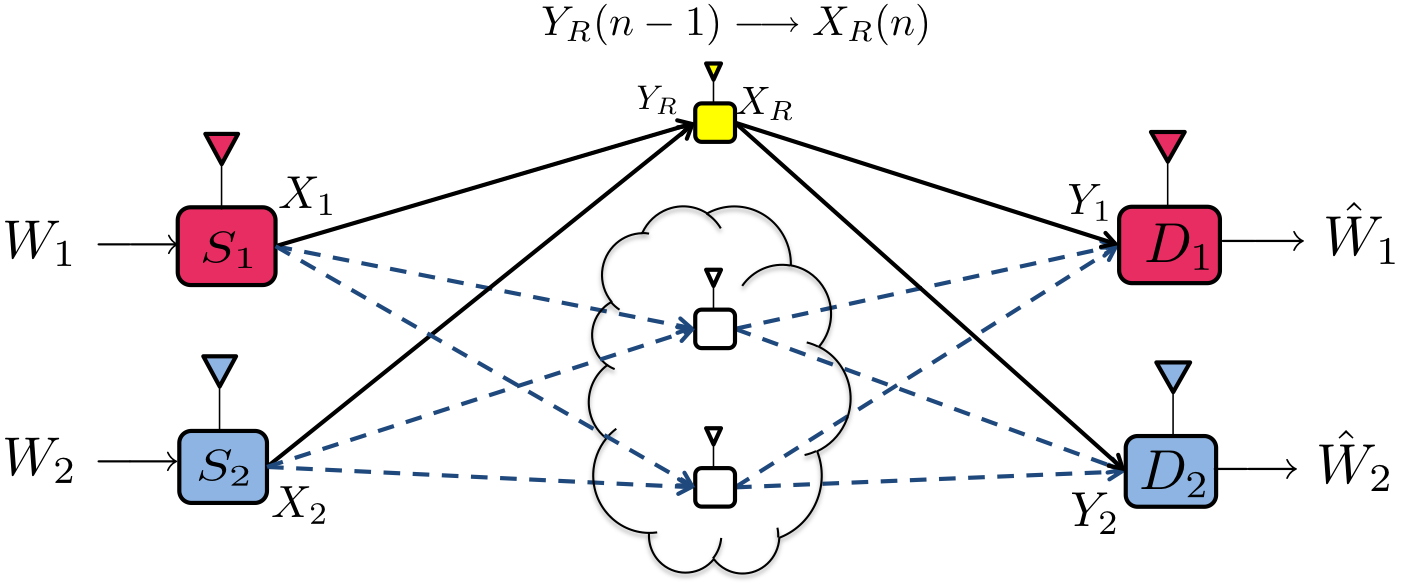}
\caption{Practical Motivation for Instantaneous Relay} \label{fig:practical}
\end{figure}

Practical  scenarios that can be modeled by the instantaneous relay setting can arise in a number of ways. Here we outline one motivating example, shown in Fig. \ref{fig:practical}. Consider a two user interference channel where communication takes place over two hops with the use of conventional relays. There are no direct links between sources and destinations, presumably because of the excessive propagation path loss, that necessitates the use of relay nodes. It is well known that in such a network,  one can completely eliminate all interference to achieve the min-cut bound $DoF=2$ with $3$ or more relay nodes, \emph{provided we have global channel knowledge and intelligent relay nodes}. However, both global channel knowledge and intelligence at all relays are not easily available in practice. A practical alternative is to choose a subset of relay nodes which are intelligent and let these relay nodes acquire the channel knowledge they need, while the remaining relay nodes remain \emph{dumb}, e.g., by simply amplifying and forwarding their received signals with no knowledge of channel conditions, nor any ability to tailor their transmissions to the channel conditions. Such a setting is shown in Fig. \ref{fig:practical}, where the highlighted relay node in yellow is the intelligent relay that acquires the channel knowledge that it needs. All other relay nodes (indicated in white) remain oblivious of the channel states and simply amplify and forward their signals with a fixed amplification factor that does not depend on channel state. Thus, the channels between sources and destinations that pass through the dumb relays can be represented as an effective channel matrix and equivalently viewed as a direct channel between the sources and destinations. It is this effective channel that is learnt by the destination nodes, in addition to the channels seen from the intelligent relay node.

\begin{figure}
\centering
\includegraphics[width=4in]{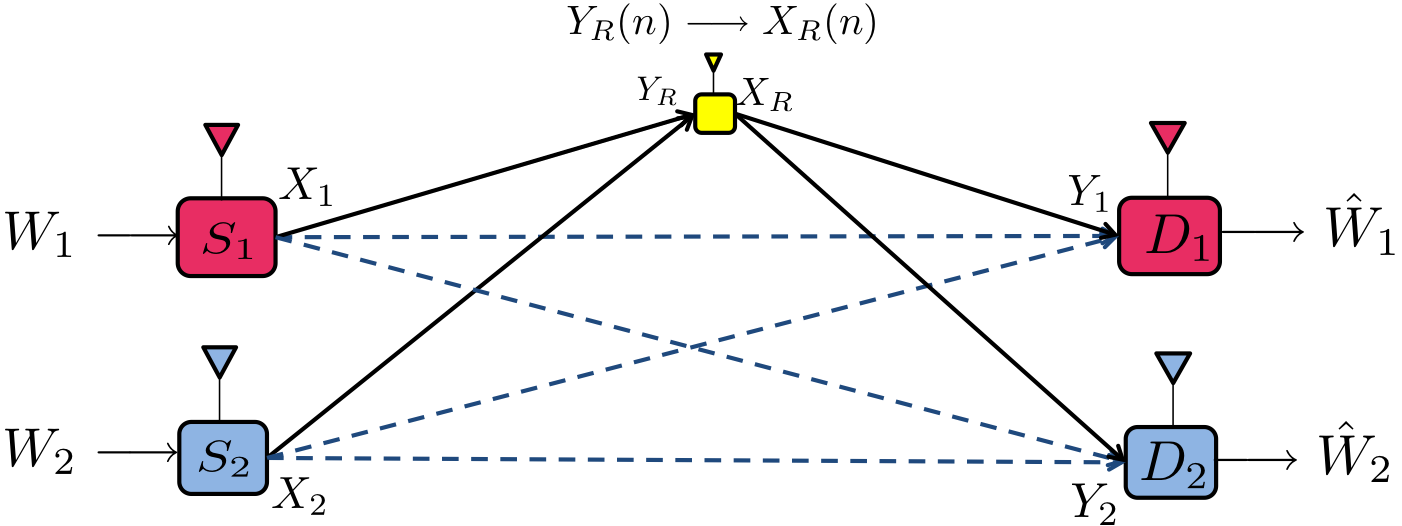}
\caption{Interference Channel with Memoryless Instantaneous  Relay} \label{fig:rwd}
\end{figure}

These practical assumptions lead us to the model shown in Fig. \ref{fig:rwd}, where the dumb relays are not explicitly shown and instead only the effective channel is retained (shown with dashed lines). Since all signals arrive at the destinations with equal delay of two hops, and the dumb relays are ignored, we correspondingly shift the time reference at the intelligent relay, so that it is modeled as an instantaneous relay. The assumption of \emph{memoryless} instantaneous relay is not essential to this work, since we focus only on achievable schemes, which will remain achievable even if the relay is allowed to have memory. Rather it is just a feature of our achievable scheme that the relays do not need to use any memory of past received symbols. Finally, we assume that the relays are full-duplex in our channel model Fig. \ref{fig:rwd}, which could again be justified as a consequence of an underlying protocol in the practical setting of Fig. \ref{fig:practical} whereby the relays transmit and receive in two orthogonal phases. Thus, the channel model of the interference channel with instantaneous relay arises naturally from a two-hop interference channel with several dumb relay nodes and one intelligent relay node.

Henceforth, we will use only the channel model illustrated in Fig. \ref{fig:rwd}, i.e., with the assumptions of instantaneous, full-duplex relays. The justification for these assumptions will not be repeated.

\subsection*{\it Contribution}

The main contribution of this paper is to show that $\frac{3}{2}$
DoF is achievable for the interference channel with the
instantaneous relay. This is remarkable in light of the result that conventional relays do not increase DoF beyond 1. Thus, there is a very significant DoF gain from the relay's ability to instantly forward the received signal from the sources to the destinations. Also surprising is that this remarkable DoF gain is achievable without any memory at the relay. The DoF gain extends to the setting where every node
is equipped with $M \geq 1$ antennas. In this case,  $\frac{3M}{2}$ DoF are achieved The result is also interesting because of the non-trivial nature of the achievable scheme, which is inspired by the idea of \emph{aligned interference neutralization} recently introduced by Gou et. al. in \cite{Gou2}.

For the multiple antenna case, $M>1$,  we
construct a  linear beamforming scheme to achieve aligned
interference neutralization. The  two sources and the relay cooperatively
construct beamforming vectors so that the interference signals
coming from different paths are eliminated at the unintended
destination. Further, we will also describe a simple
modification of the beamforming scheme that can achieve SNR gain for
desired data signals while maintaining $\frac{3M}{2}$ DoF. The
modified scheme  designs the beamforming vectors in such a way
that not only are the interference signals coming from different paths eliminated at the unintended destination, but also the desired
signals coming from different paths are coherently combined at the
desired destinations.

For single antenna case, i.e., $M=1$, we translate the proposed linear beamforming scheme  into the rational dimensions framework introduced in
\cite{Motahari1} and \cite{Motahari2}. Using this framework, we
will also show that $\frac{3}{2}$ DoF are achieved for the
interference channel with the instantaneous relay when the channel coefficient values remain fixed.

The rest of the paper is organized as follows. In Section II, the
system model is described and the key concept of proposed scheme
is precisely explained using a simple example in Section III. In
Section IV we derive the lower bound of DoF for the interference
channel with the instantaneous relay, assuming all nodes have
multiple antennas. Using the rational dimension framework, we also
derive the lower bound of DoF for a single antenna system in
Section V. The paper is concluded in Section VI with discussions
for the role of a relay in the interference channel.

\section{System model}
As shown in Fig. \ref{fig:rwd} our channel model consists of two sources, two destinations, and an intermediate
relay. In this channel, source  $S_i$, $i=1,2$,  wishes to send an
independent message $W^{[i]}$ to its respective destination $D_i$. Global channel knowledge is assumed at all 5 nodes, with the notable exception that the source nodes need not learn the channel coefficients for the direct channels to the destination nodes (i.e., the dashed links in Fig. \ref{fig:rwd}, i.e., the source nodes only need to learn their channels to the relay node. The relay operates in full-duplex mode and it
instantaneously forwards the currently received data symbols from
the sources to the destinations. For
a multiple antenna system, input-output relationships of the
channel are given by
\begin{eqnarray}
{\bf y}^{[R]} \!\!\!\!\!&=&\!\!\!\!\! \sum_{i=1}^{2}{\bf H}^{[R,i]}{\bf x}^{[i]} \!+\! {\bf n}^{[R]} \nonumber \\
{\bf y}^{[1]} \!\!\!\!\!&=&\!\!\!\!\!  \sum_{i=1}^{2}{\bf H}^{[1,i]}{\bf x}^{[i]}+{\bf H}^{[1,R]}{\bf x}^{[R]}\!+\! {\bf n}^{[1]} \nonumber \\
{\bf y}^{[2]} \!\!\!\!\!&=&\!\!\!\!\! \sum_{i=1}^{2}{\bf
H}^{[2,i]}{\bf x}^{[i]}+{\bf H}^{[2,R]}{\bf x}^{[R]}\!+\! {\bf
n}^{[2]},
\end{eqnarray}
where ${\bf y}^{[j]}$, $j\in\{1,2,R\}$ denotes the channel output
vector with size of $M\times 1$ at Receiver $j$, ${\bf
H}^{[j,i]}$ is the $M\times M$ channel matrix  from
the sender $i$ to the receiver $j$ where $i,j\in\{1,2,R\}$, and
${\bf n}^{[j]}$ $j\in\{1,2,R\}$ represents a noise signal vector
with size of $M\times 1$ at the destination $j$ whose elements are
random variables drawn from an independent and identically
distributed (i.i.d.) zero mean complex Gaussian distribution
with unit variance, i.e., $\mathcal{NC}(0,1)$. In addition, we
assume that all elements of each channel matrix are drawn from
$\mathcal{NC}(0,1)$. The average power constraint of the
transmitted signals by the sender $i$, ${\bf x}^{[i]}$, is given
by
\begin{eqnarray}
\mathbb{E}\left[\textmd{Tr}\left({\bf x}^{[i]}{{\bf
x}^{[i]}}^{H}\right)\right] \leq P, \quad i\in\{1,2,R\}.
\end{eqnarray}

For codewords spanning $n$ channel uses, a rate
$R^{[i]}(P)=\frac{\log|W^{[i]}(P)|}{n}$ is achievable if the
probability of error for the message $W^{[i]}$ approaches zero as
$n\rightarrow \infty$. The DoF characterizing the high SNR
behavior of the achievable rate is defined as
\begin{eqnarray}
d^{[i]}  =\lim_{P\rightarrow \infty} \frac{R^{[i]}(P)}{\log(P)},
\quad i\in\{1,2\}.
\end{eqnarray}

\section{Proposed scheme}
For ease of exposition, we first explain our proposed scheme with a simple example
when $M=4$.

\begin{figure}
\centering
\includegraphics[width=4.0in]{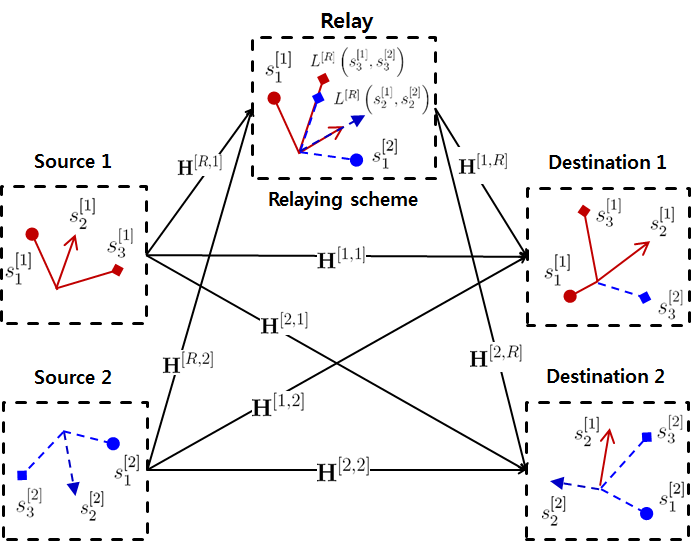}
\caption{The proposed scheme for the interference channel with the
instantaneous relay when $M=4$.} \label{fig_sim}
\end{figure}

\subsection{Transmission scheme at two source nodes}
As shown in Fig. \ref{fig_sim}, Source  $S_i$ wishes to deliver three
independent data symbols $s^{[i]}_{1}$, $s^{[i]}_{2}$, and
$s^{[i]}_{3}$ by using linear beamforming vectors ${\bf
v}^{[i]}_{1}$, ${\bf v}^{[i]}_{2}$, and ${\bf v}^{[i]}_{3}$ to the
corresponding destination  $D_i$, where $i\in\{1,2\}$. Since
each source transmits three independent data symbols, each
destination and the receiver of the relay node see six competing flows of
information. If each receiver has $M\geq6$ antennas, it is easily
seen that without the help of the relay, each destination node
can get three desired data symbols by removing the interfering
symbols. However, when we only have $M=4$ antennas, the two destinations cannot
by default resolve the 3 desired symbols from the 3 interfering symbols when confined to an overall four dimensional space.  To do
this, the two source nodes cooperatively design transmit beamforming
vectors so that the relay sees the following four data symbols in
a four dimensional space: ${s}^{[1]}_{1}$, ${s}^{[2]}_{1}$,
$L^{[R]}\left({s}^{[1]}_{2},{s}^{[2]}_{2}\right)$, and
$L^{[R]}\left({s}^{[1]}_{3},{s}^{[2]}_{3}\right)$ where
$L^{[j]}(a,b)$ denotes a linear combination of $a$ and $b$ at the
$j$-th receiver. To achieves these goals, the beamforming vectors
of two source nodes are picked as
\begin{eqnarray}
\textrm{span}\left({\bf H}^{[R,1]}{\bf v}^{[1]}_{2}\right)&=&\textrm{span}\left({\bf H}^{[R,2]}{\bf v}^{[2]}_{2}\right) \nonumber \\
\textrm{span}\left({\bf H}^{[R,1]}{\bf
v}^{[1]}_{3}\right)&=&\textrm{span}\left({\bf H}^{[R,2]}{\bf
v}^{[2]}_{3}\right),
\end{eqnarray}
where \textrm{span}$\left({\bf A}\right)$ stands for the vector
space spanned by the column vectors of matrix ${\bf A}$. Note that if the source nodes know only their own channels to the destination node, then it is also possible to achieve the same alignments, by having the relay choose offline the dimensions along which the alignments should occur. Since this choice is made offline, it is known to the source nodes, who can then design their beams to reach the relay in the correct dimension, without needing to learn the other source's channel to the relay node.

\subsection{Relaying strategy}
Note that each destination sees six independent data symbols from
two sources nodes while it has four antennas. Therefore, the relay
transmission strategy is to forward a signal so that it
simultaneously removes at least two interference data symbols at
each destination. We know that the relay estimates four data
symbols from two source nodes, which are $s^{[1]}_{1}$,
${s}^{[2]}_{1}$,
$L^{[R]}\left({s}^{[1]}_{2},{s}^{[2]}_{2}\right)$, and
$L^{[R]}\left({s}^{[1]}_{3},{s}^{[2]}_{3}\right)$. Using the
estimated data symbols, we first generate transmit data symbols
sent by the relay, which are
\begin{eqnarray}
{s}^{[R]}_1&=&{s}^{[1]}_1, \quad\quad\quad\quad\quad\quad  {s}^{[R]}_2 = {s}^{[2]}_1, \nonumber \\
{s}^{[R]}_3&=&L^{[R]}\left({s}^{[1]}_{2},{s}^{[2]}_{2}\right),
\quad {s}^{[R]}_4
=L^{[R]}\left({s}^{[1]}_{3},{s}^{[2]}_{3}\right).
\end{eqnarray}

 As a next step, we also design the relay beamforming
vectors ${\bf v}^{[R]}_i$ for carrying ${s}^{[R]}_i$,
$i\in\{1,2,3,4\}$. The proposed relay beamforming strategy is as
follows:
\begin{itemize}
    \item Relay picks ${\bf v}^{[R]}_1$ to cancel ${s}^{[1]}_1$ at Destination 2.
    \item Relay picks ${\bf v}^{[R]}_2$ to cancel ${s}^{[2]}_1$ at Destination 1.
    \item Relay picks ${\bf v}^{[R]}_3$ to cancel ${s}^{[2]}_2$ at Destination 1.
    \item Relay picks ${\bf v}^{[R]}_4$ to cancel ${s}^{[1]}_3$ at Destination 2.
\end{itemize}
To accomplish this goal, we choose the relay beamforming vectors
${\bf v}^{[R]}_i$ as
\begin{eqnarray}
{\bf H}^{[2,R]}{\bf v}^{[R]}_1&=&-{\bf H}^{[2,1]}{\bf
v}^{[1]}_{1}, \quad
{\bf H}^{[1,R]}{\bf v}^{[R]}_2=-{\bf H}^{[1,2]}{\bf v}^{[2]}_{1} \nonumber \\
{\bf H}^{[1,R]}{\bf v}^{[R]}_3&=&-{\bf H}^{[1,2]}{\bf
v}^{[2]}_{2}, \quad {\bf H}^{[2,R]}{\bf v}^{[R]}_4=-{\bf
H}^{[2,1]}{\bf v}^{[1]}_{3}.
\end{eqnarray}

\subsection{Decoding at destinations}
Now we need to check whether each destination can decode three
desired data symbols. From the relay transmission, two of the
undesired signals are neutralized at each destination, leaving
only four signals which can be resolved in a four dimensional
space. The remaining data symbols at each destination are
\begin{itemize}
    \item Destination 1 observes four signals: ${s}^{[1]}_1$,
    ${s}^{[1]}_2$, ${s}^{[1]}_3$, and ${s}^{[2]}_3$.
    \item Destination 2 observes four signals: ${s}^{[2]}_1$,
    ${s}^{[2]}_2$, ${s}^{[2]}_3$, and ${s}^{[1]}_2$.
\end{itemize}
As a result, each destination can decode three independent data
symbols and thus 6 DoF are achieved when $M=4$. This is interesting
because it is higher than the 4 DoF that can be
achieved for the two-user MIMO interference channel without
involving the relay node \cite{JafarIC}.

\section{Achievable DoF for Multiple Antenna Systems}

The following theorem is the main result of this paper.

\textbf{\emph{Theorem 1}}: \label{Theorem1} In a complex MIMO
Gaussian interference channel with an instantaneous relay,
$\frac{3M}{2}$ DoF is achievable for $M>1$.

Proof:) The achievability proof is shown by aligned interference
neutralization with interference-forward relaying scheme. Here, we
will show $(d^{[1]},d^{[2]})=(\frac{3M}{4},\frac{3M}{4})$ when $M$
is multiple of 4 for simplicity. For arbitrary $M$, we can
straightforwardly show that the same DoF can be achieved using 4
time symbol extension technique as shown in the both the MIMO X
channel \cite{JafarX} and the MIMO Y channel \cite{namyoon1}. However, note that the symbol extension based achievability schemes, needed when $M$ is not a multiple of $4$,  do require memory in the relay processing, over 4 consecutive symbols. This requirement can be reduced by exploiting asymmetric complex signaling, in which case only two symbol extensions would suffice.

\subsubsection*{Sources}
Let us start by splitting each transmit message at source nodes
into three submessages, i,e.,
$W^{[1]}=\{W^{[1]}_{1},W^{[1]}_{2},W^{[1]}_{3}\}$ and
$W^{[2]}=\{W^{[2]}_{1},W^{[2]}_{2},W^{[2]}_{3}\}$. Each submessage
$W^{[i]}_{j}$ is encoded into $r=\frac{M}{4}$ transmit symbols,
i.e., $s^{[i]}_{j,1},s^{[i]}_{j,2},\ldots,s^{[i]}_{j,r}$, using a
Gaussian codebook to be transmitted over $n$ channel uses where $i\in\{1,2\}$ and
$j\in\{1,2,3\}$. After encoding, the two source nodes send data symbols
$s^{[i]}_{j,k}$ along beamforming vectors ${\bf
v}^{[i]}_{j,k}$. Thus, signals sent by the two source nodes are
given by
\begin{eqnarray}
{\bf x}^{[i]}&=&\sum_{k=1}^{r}\sum_{j=1}^{3}{\bf
v}^{[i]}_{j,k}s^{[i]}_{j,k} \nonumber \\
&=&\sum_{j=1}^{3}{\bf V}^{[i]}_{j}{\bf s}^{[i]}_{j}, \quad
i\in\{1,2\},
\end{eqnarray}
where ${\bf V}^{[i]}_{j}=\left[{\bf v}^{[i]}_{j,1},{\bf
v}^{[i]}_{j,2},\ldots, {\bf v}^{[i]}_{j,r}\right]$ and ${\bf
s}^{[i]}_{j}=\left[{s}^{[i]}_{j,1},{s}^{[i]}_{j,2},\ldots,
{s}^{[i]}_{j,r}\right]^{T}$, $j\in\{1,2,3\}$.

The transmit beamforming strategy of  the two source nodes is to select
the $4r$ beamforming directions so that the following alignment
conditions are satisfied,
\begin{eqnarray}
\textrm{span}\left({\bf H}^{[R,1]}{\bf V}^{[1]}_{2}\right)&=&\textrm{span}\left({\bf H}^{[R,2]}{\bf V}^{[2]}_{2}\right) \nonumber \\
\textrm{span}\left({\bf H}^{[R,1]}{\bf
V}^{[1]}_{3}\right)&=&\textrm{span}\left({\bf H}^{[R,2]}{\bf
V}^{[2]}_{3}\right).
\end{eqnarray}


\subsubsection*{Relay}
The relaying scheme involves two steps:

1) estimate $4r$ symbols,
${\hat{s}}^{[1]}_{1,k}$, ${\hat{s}}^{[2]}_{1,k}$,
$L^{[R]}_k\left({\hat{s}}^{[1]}_{2,k},{\hat{s}}^{[2]}_{2,k}\right)$,
and
$L^{[R]}_k\left({\hat{s}}^{[1]}_{3,k},{\hat{s}}^{[2]}_{3,k}\right)$
where $k\in\{1,2,\ldots,r\}$

2) forward the estimated symbols in
such a way that interference signals are neutralized at each
destination. From (8), the received signal at the relay is given
by
\begin{eqnarray}
{\bf y}^{[R]} \!=\! \left[%
\begin{array}{cccc}
 {\bf H}^{[R,1]}{\bf V}^{[1]}_{1} & {\bf H}^{[R,2]}{\bf V}^{[2]}_{1} & {\bf H}^{[R,2]}{\bf
V}^{[2]}_{2} & {\bf H}^{[R,1]}{\bf V}^{[1]}_{3}\\
\end{array}%
\right]\left[\!\!%
\begin{array}{c}
  {\bf s}^{[1]}_{1} \\
  {\bf s}^{[2]}_{1} \\
  \Lambda^{[R,1]}_2{\bf s}^{[1]}_{2} \!\!+\!\! {\bf s}^{[2]}_{2}\\
  {\bf s}^{[1]}_{3} \!\!+\!\! \Lambda^{[R,2]}_3{\bf s}^{[2]}_{3} \\
\end{array}%
\!\!\right] \!+\! {\bf n}^{[R]},
\end{eqnarray}
where $\Lambda^{[R,1]}_2$ and $\Lambda^{[R,2]}_3$ are diagonal
matrices with size of $r\times r$. In addition, the $l$-th
diagonal elements of $\Lambda^{[R,1]}_2$ and $\Lambda^{[R,2]}_3$
are defined as $\Lambda^{[R,1]}_2(l,l)=\|{\bf H}^{[R,1]}{\bf
v}^{[1]}_{2,l}\|/\|{\bf H}^{[R,2]}{\bf v}^{[2]}_{2,l}\|$ and
$\Lambda^{[R,2]}_3(l,l)=\|{\bf H}^{[R,2]}{\bf
v}^{[2]}_{3,l}\|/\|{\bf H}^{[R,1]}{\bf v}^{[1]}_{3,l}\|$,
respectively. Since ${\bf H}^{[R,1]}{\bf V}^{[1]}_{1}$, ${\bf
H}^{[R,2]}{\bf V}^{[2]}_{1}$, ${\bf H}^{[R,1]}{\bf V}^{[1]}_{2}$,
and ${\bf H}^{[R,1]}{\bf V}^{[1]}_{3}$ are linearly independent
with probability one, the relay can reliably estimates $4r$
symbols, ${\hat{s}}^{[1]}_{1,k}$, ${\hat{s}}^{[2]}_{1,k}$,
$L^{[R]}_k\left({\hat{s}}^{[1]}_{2,k},{\hat{s}}^{[2]}_{2,k}\right)$,
and
$L^{[R]}_k\left({\hat{s}}^{[1]}_{3,k},{\hat{s}}^{[2]}_{3,k}\right)$
where $k\in\{1,2,\ldots,r\}$ if we consider high \textsf{SNR}
regime.

After estimating transmit symbols, the relay sends
${s}^{[R]}_{j,k}$ along a beamforming vector ${\bf
v}^{[R]}_{j,k}$ which results in the transmitted vector
\begin{eqnarray}
{\bf x}^{[R]}&=&\sum_{j=1}^{4}\sum_{k=1}^{r}{\bf
v}^{[R]}_{j,k}{s}^{[R]}_{j,k} \nonumber \\
&=&\sum_{j=1}^{4}{\bf V}^{[R]}_j{\bf s}^{[R]}_j,
\end{eqnarray}
where $ {s}^{[R]}_{1,k}={\hat{s}}^{[1]}_{1,k}$,
${s}^{[R]}_{2,k}={\hat{s}}^{[2]}_{1,k}$, $
{s}^{[R]}_{3,k}=L^{[R]}_k\left({\hat{s}}^{[1]}_{2,k},{\hat{s}}^{[2]}_{2,k}\right)$,
and
${s}^{[R]}_{4,k}=L^{[R]}_k\left({\hat{s}}^{[1]}_{3,k},{\hat{s}}^{[2]}_{3,k}\right)$
for $k\in\{1,2,\ldots,r\}$.

The relay beamforming strategy aims at removing the interference
signals coming from the undesired source node. Specifically,
Destination 1 receives $3r$ interference signals sent by Source 2,
${\bf H}^{[R,2]}{\bf V}^{[2]}_{j}$ where $j\in\{1,2,3\}$.
Similarly, Destination 2 also receives $3r$ interference signals
sent by Source 1, ${\bf H}^{[R,1]}{\bf V}^{[1]}_{j}$ where
$j\in\{1,2,3\}$. Therefore, the relay helps both destination nodes
 eliminate enough of those interference signals to  enable
each destination to decode the desired signals. To do this, the
relay beamforming vectors ${\bf v}^{[R]}_{j,k}$ are designed as
\begin{eqnarray}
{\bf v}^{[R]}_{1,k}&=& -{{\bf H}^{[2,R]}}^{-1}{\bf H}^{[2,1]}{\bf v}^{[1]}_{1,k} \nonumber \\
{\bf v}^{[R]}_{2,k}&=& -{{\bf H}^{[1,R]}}^{-1}{\bf H}^{[1,2]}{\bf v}^{[2]}_{1,k} \nonumber \\
{\bf v}^{[R]}_{3,k}&=& -{{\bf H}^{[1,R]}}^{-1}{\bf H}^{[1,2]}{\bf v}^{[2]}_{2,k} \nonumber \\
{\bf v}^{[R]}_{4,k} &=& -{{\bf H}^{[2,R]}}^{-1}{\bf H}^{[2,1]}{\bf
v}^{[1]}_{3,k}.
\end{eqnarray}
Note that ${\bf v}^{[2]}_{2,k}$ and ${\bf v}^{[1]}_{3,k}$ were
already picked when two source nodes transmit. In addition, we can
choose ${\bf v}^{[1]}_{1,k}$ and ${\bf v}^{[2]}_{1,k}$ randomly
for $k\in\{1,2,\ldots,r\}$.

\subsubsection*{Destinations}
So far, we explain the transmit beamforming strategy at each
source node and the relaying scheme. Now, we need to check the
decodablity of the desired symbols at each destination node. Let
us consider Destination 1, which wants to decode three independent
symbol vectors ${\bf s}^{[1]}_{j}$, $j\in\{1,2,3\}$. The received
signal at Destination 1 is given by
\begin{eqnarray}
{\bf y}^{[1]}&\!\!\!=\!\!\!&\sum_{i=1}^{2}{\bf H}^{[1i]}{\bf x}^{[i]}+{\bf H}^{[1,R]}{\bf x}^{[R]}\!+\! {\bf n}^{[1]} \nonumber \\
&\stackrel{(a)}{=}& \left({\bf H}^{[1,1]}{\bf V}^{[1]}_{1}+{\bf
H}^{[1,R]}{\bf V}^{[R]}_{1}\right){\bf s}^{[1]}_{1} + \left({\bf
H}^{[1,1]}{\bf V}^{[1]}_{2}+{\bf H}^{[1,R]}{\bf
V}^{[R]}_{3}\Lambda^{[R,1]}_2\right){\bf s}^{[1]}_{2} \nonumber \\
&&+ \left({\bf H}^{[1,1]}{\bf V}^{[1]}_{3}+{\bf H}^{[1,R]}{\bf
V}^{[R]}_{4}\right){\bf s}^{[1]}_{3} +  \left({\bf H}^{[1,2]}{\bf
V}^{[2]}_{3}+{\bf H}^{[1,R]}{\bf
V}^{[R]}_{4}\Lambda^{[R,2]}_3\right){\bf s}^{[2]}_{3}\!+\! {\bf
n}^{[1]} \nonumber \\
&=&\left[\!\!\!
\begin{array}{cccc}
  {\bf \bar{H}}^{[1,1]}_1 & {\bf \bar{H}}^{[1,1]}_2 & {\bf \bar{H}}^{[1,1]}_3 & {\bf \bar{H}}^{[1,2]}_3 \\
\end{array}%
\!\!\!\right]\!\!\left[\!\!\!
\begin{array}{c}
  {\bf s}^{[1]}_{1} \\
  {\bf s}^{[1]}_{2} \\
  {\bf s}^{[1]}_{3} \\
  {\bf s}^{[2]}_{3} \\
\end{array}%
\!\!\!\right]\!+\! {\bf n}^{[1]},
\end{eqnarray}
where $(a)$ comes from the conditions in (11) and ${\bf
\bar{H}}^{[1,i]}_j$ denotes effective channel carrying the symbol
vector ${\bf s}^{[i]}_{j}$. From (12) it easily can be seen that
$3r$ desired symbols, ${s}^{[1]}_{j,k}$ for $j\in\{1,2,3\}$ and
$k\in\{1,2,\ldots,r\}$, can be decoded by applying zero-forcing
since ${\bf \bar{H}}^{[1,1]}_1$, ${\bf \bar{H}}^{[1,1]}_2$, ${\bf
\bar{H}}^{[1,1]}_3$, and ${\bf \bar{H}}^{[1,2]}_3$ are linearly
independent. As a result, Destination 1 can achieve $d_1=3r$ DoF.

In the same way, Destination 2 also can decode three independent
signal vectors ${\bf s}^{[2]}_{j}$, $j\in\{1,2,3\}$. The received
signal at Destination 2 is given by
\begin{eqnarray}
{\bf y}^{[2]}&\!\!\!=\!\!\!&\sum_{i=1}^{2}{\bf H}^{[2i]}{\bf x}^{[i]}+{\bf H}^{[2,R]}{\bf x}^{[R]}\!+\! {\bf n}^{[2]} \nonumber \\
&\stackrel{(a)}{=}& \left({\bf H}^{[2,2]}{\bf V}^{[2]}_{1}+{\bf
H}^{[2,R]}{\bf V}^{[R]}_{2}\right){\bf s}^{[2]}_{1} + \left({\bf
H}^{[2,2]}{\bf V}^{[2]}_{2}+{\bf H}^{[2,R]}{\bf
V}^{[R]}_{3}\right){\bf s}^{[2]}_{2} \nonumber \\
&&+ \left({\bf H}^{[2,2]}{\bf V}^{[2]}_{3}+{\bf H}^{[2,R]}{\bf
V}^{[R]}_{4}\Lambda^{[R,2]}_3\right){\bf s}^{[2]}_{3} + \left({\bf
H}^{[2,1]}{\bf V}^{[1]}_{2}+{\bf H}^{[2,R]}{\bf
V}^{[R]}_{3}\Lambda^{[R,1]}_2\right){\bf s}^{[1]}_{2}\!+\! {\bf
n}^{[2]} \nonumber \\
&=&\left[\!\!\!
\begin{array}{cccc}
  {\bf \bar{H}}^{[2,2]}_1 & {\bf \bar{H}}^{[2,2]}_2 & {\bf \bar{H}}^{[2,2]}_3 & {\bf \bar{H}}^{[2,1]}_2 \\
\end{array}%
\!\!\!\right]\!\!\left[\!\!\!
\begin{array}{c}
  {\bf s}^{[2]}_{1} \\
  {\bf s}^{[2]}_{2} \\
  {\bf s}^{[2]}_{3} \\
  {\bf s}^{[1]}_{2} \\
\end{array}%
\!\!\!\right]\!+\! {\bf n}^{[2]},
\end{eqnarray}
where $(a)$ also comes from the interference neutralization
conditions in (11) and ${\bf \bar{H}}^{[2,i]}_j$ denotes effective
channel matrix carrying the symbol vector ${\bf s}^{[i]}_{j}$.
Since ${\bf \bar{H}}^{[2,2]}_1$, ${\bf \bar{H}}^{[2,2]}_2$, ${\bf
\bar{H}}^{[2,2]}_3$, and ${\bf \bar{H}}^{[2,1]}_2$ are linearly
independent, Destination 2 also can achieve $d_2=3r$ DoF.
Consequently, $(d^{[1]},d^{[2]})=(\frac{3M}{4},\frac{3M}{4})$ is
achieved when $r=\frac{M}{4}$.

\textbf{\emph{Remark 1}}: The proposed beamforming scheme for two
source nodes and the relay focuses on attaining DoF gain. However,
we can also design beamforming vectors of two sources and the
relay to obtain diversity gain while keeping $\frac{3M}{2}$ DoF.
For example, Destination 1 receives the desired symbol vector
${\bf s}^{[1]}_1$ through the effective channel ${\bf
\bar{H}}^{[1,1]}_1={\bf H}^{[1,1]}{\bf V}^{[1]}_{1}+{\bf
H}^{[1,R]}{\bf V}^{[R]}_{1}$. Note that ${\bf V}^{[R]}_{1}$ was
designed for neutralizing the interference symbol ${\bf
s}^{[1]}_{1}$ at Destination 2. However, Destination 1 can
overhear the relaying signals ${\bf H}^{[1,R]}{\bf
V}^{[R]}_{1}{\bf s}^{[1]}_{1}$. Therefore, to increase diversity
gain, the beamforming vectors of Source 1 is constructed as
\begin{eqnarray}
\max_{{\bf v}^{[1]}_{1,k}}\|{\bf H}^{[1,1]}{\bf
v}^{[1]}_{1,k}+{\bf H}^{[1,R]}{\bf v}^{[R]}_{1,k}\|_{2}, \quad
k\in\{1,2,\ldots,r\}.
\end{eqnarray}
Similarly, Source 2 can pick beamforming vectors ${\bf
v}^{[2]}_{1,k}$ as
\begin{eqnarray}
\max_{{\bf v}^{[2]}_{1,k}}\|{\bf H}^{[2,2]}{\bf
v}^{[2]}_{1,k}+{\bf H}^{[2,R]}{\bf v}^{[R]}_{2,k}\|_{2}, \quad
k\in\{1,2,\ldots,r\}.
\end{eqnarray}
From (14) and (15), we can increase effective channel gains by
coherently combining the desired symbols coming from two different
pathes. \emph{As a remark, the transmitted signals sent by the
relay can not only neutralize interference signals at the
unintended destination (yielding DoF gain), but also can increase
effective channel gains for desired signals at the intended
destination (yielding SNR gain)}. However, note that in order to do so, the source nodes need channel knowledge for the direct channels to the destination nodes.

\section{Single antenna case}
In this section, we will show that $\frac{3}{2}$ DoF is achieved
for the interference channel with an instantaneous relay when
$M=1$, i.e., single antenna case. To show this result we use the
rational dimension framework introduced in \cite{Motahari1} and
\cite{Motahari2} for the alignment technique.
\subsection{System model for $M=1$}
For a single antenna system, the channel's input-output
relationship is given by
\begin{eqnarray}
{y}^{[R]} \!\!\!\!\!&=&\!\!\!\!\! \sum_{i=1}^{2}{h}^{[R,i]}{x}^{[i]} \!+\! {n}^{[R]} \nonumber \\
{y}^{[1]} \!\!\!\!\!&=&\!\!\!\!\!  \sum_{i=1}^{2}{h}^{[1,i]}{x}^{[i]}+{h}^{[1,R]}{x}^{[R]}\!+\! {n}^{[1]} \nonumber \\
{y}^{[2]} \!\!\!\!\!&=&\!\!\!\!\!
\sum_{i=1}^{2}{h}^{[2,i]}{x}^{[i]}+{h}^{[2,R]}{x}^{[R]}\!+\! {
n}^{[2]},
\end{eqnarray}
where ${y}^{[j]}$, $j\in\{1,2,R\}$ denotes the channel output
signal at the receiver $j$, ${h}^{[j,i]}$ denotes channel gain
coefficient from the sender $i$ to the receiver $j$ where
$i,j\in\{1,2,R\}$. It is assumed that all channel gains are real
and time invariant. In (14), ${n}^{[j]}$ $j\in\{1,2,R\}$
represents a AWGN noise, i.e., $\mathcal{NC}(0,1)$ at the receiver
$j$. In addition, ${x}^{[i]}$, $i\in\{1,2,R\}$ are the transmitted
signal by the sender $i$, which has an average power constraint as
\begin{eqnarray}
\mathbb{E}\left[{x^{[i]}}^2\right] \leq P, \quad i\in\{1,2,R\}.
\end{eqnarray}
Unlike the multiple antenna case, we consider real channel
coefficient. Therefore, the DoF is defined as
\begin{eqnarray}
d_i =\lim_{P\rightarrow \infty} \frac{R_i(P)}{\frac{1}{2}\log(P)},
\quad i\in\{1,2\}.
\end{eqnarray}

\begin{figure}
\centering
\includegraphics[width=4.5in]{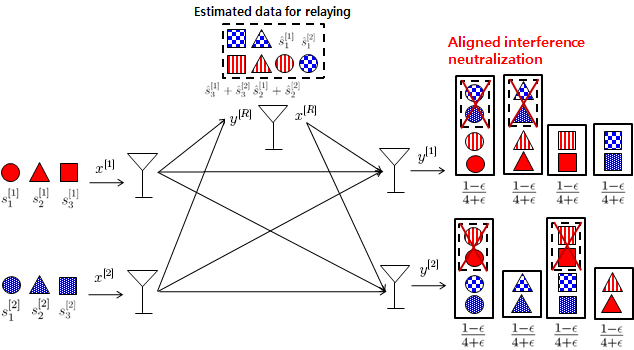}
\caption{Proposed scheme for a single antenna case.}
\label{fig_simreal}
\end{figure}

\subsection{Achievable DoF when $M=1$}

\textbf{\emph{(Theorem 2)}} \label{Theorem2} For a interference
relay channel with the instantaneous relay, $\frac{3}{2}$ DoF is
achievable when all nodes have a single antenna, i.e., $M=1$.

Proof:) \subsubsection{Sources} As illustrated in Fig. 2, each
ource, $S_i$, $i=\{1,2\}$ divides the message $W^{[i]}$ into three
submessages $W^{[i]}_{1}$, $W^{[i]}_{2}$, and $W^{[i]}_{3}$.
Submessage $W^{[i]}_{j}$ is encoded to transmit data symbols
employing a codebook consisted of the codeword having length $n$,
denoted by $\{s^{[i]}_j(1),\ldots,s^{[i]}_j(n)\}$ for
$j\in\{1,2,3\}$. Let define transmit constellation
$\mathcal{C}_{S}=[-Q,Q]_\mathbb{Z}=\{-Q,-Q+1,\ldots,Q-1,Q\}$,
where $Q$ is the integer. Using this constellation, the
transmitted data symbol is uniformly selected on
$\mathcal{C}_{S}$. The transmitted signal at $S_i$ is the linear
combination of three integer data symbols $s^{[i]}_1$,
$s^{[i]}_2$, and $s^{[i]}_3$, which is denoted by
\begin{eqnarray}
{x}^{[i]}&=&A\sum_{j=1}^{3}{v}^{[i]}_{j}s^{[i]}_{j},
\end{eqnarray}
where $A$ is a power normalizing constant to meet the transmit
power constraint and ${v}^{[i]}_{j}$ $j\in\{1,2,3\}$ is real
coefficient determining the transmit direction of the integer data
symbol $s^{[i]}_{j}$ in the rational domain. The transmit power of
$S_i$ is bounded as
\begin{eqnarray}
E\left[{{x}^{[i]}}^2\right]&=&A^2\underbrace{\sum_{j=1}^{3}{{v}^{[i]}_{j}}^2}_{{\eta^{[i]}}^2}E\left[{{s}^{[i]}_j}^2\right]\leq
A^2{\eta^{[i]}}^2Q^2 \leq P,
\end{eqnarray}
For satisfying the transmit power constraint, we pick $A$ as
\begin{eqnarray}
A = \frac{\eta P^{\frac{1}{2}}}{Q},
\end{eqnarray}
where $\eta=\frac{1}{\min\{\eta^{[1]},\eta^{[2]}\}}$. Similar to
linear beamforming scheme in the previous section, as depicted in
Fig. 2, we design real coefficients ${v}^{[i]}_{j}$,
$i,j\in\{1,2,3\}$ in order to satisfy the alignment conditions at
the both destinations and the relay, which are
\begin{eqnarray}
{h}^{[R,1]}{v}^{[1]}_{2}&=&{h}^{[R,2]}{v}^{[2]}_{2}, \nonumber \\
{h}^{[R,1]}{v}^{[1]}_{3}&=&{h}^{[R,2]}{v}^{[2]}_{3}.
\end{eqnarray}
Note that the solutions of ${v}^{[i]}_{j}$ for $i\in\{1,2\}$ and
$j\in\{2,3\}$ are obtained with probability one.

\subsubsection{Relay}
The role of the relay is to forward the interference signals so
that the sum of the interference signal is canceled out at the
unintended receiver. To do this, the relay first reliably estimate
four data symbols ${s}^{[1]}_{1}$, ${s}^{[2]}_{1}$,
${s}^{[1]}_{2}+{s}^{[2]}_{2}$, and ${s}^{[1]}_{3}+{s}^{[2]}_{3}$
for sending to both destinations. The received signal at the relay
$R$ is
\begin{eqnarray}
y^{[R]}&=&{h}^{[R,1]}x^{[1]}+{h}^{[R,2]}x^{[2]}+n^{[R]} \nonumber \\
&=&{h}^{[R,1]}A\sum_{j=1}^{3}{v}^{[1]}_{j}s^{[1]}_{j}+{h}^{[R,2]}A\sum_{j=1}^{3}{v}^{[2]}_{j}s^{[2]}_{j}+n^{[R]} \nonumber \\
&\stackrel{(a)}{=}&A{h}^{[R,1]}{v}^{[1]}_{1}s^{[1]}_{1}\!+\!
A{h}^{[R,2]}{v}^{[2]}_{1}s^{[2]}_{1}\!+\!
A{h}^{[R,1]}{v}^{[1]}_{2}\left(s^{[1]}_{2}+s^{[2]}_{2}\right)
\!+\!A{h}^{[R,1]}{v}^{[1]}_{3}\left(s^{[1]}_{3}+s^{[2]}_{3}\right)\!+\!n^{[R]},
\end{eqnarray}
where $(a)$ comes from the alignment condition (22). Note that
$s^{[1]}_{2}+s^{[2]}_{2}$ and $s^{[1]}_{3}+s^{[2]}_{3}$ are
elements of a new integer set $[-2Q,2Q]_\mathbb{Z}$. Therefore, if
we ignore the noise, the received signal at the relay is a point
of the following received signal constellation $\mathcal{C}_R$,
which is
\begin{eqnarray}
\mathcal{C}_R=\left\{A{h}^{[R,1]}{v}^{[1]}_{1}s^{[1]}_{1}\!+\!
A{h}^{[R,2]}{v}^{[2]}_{1}s^{[2]}_{1}\!+\!
A{h}^{[R,1]}{v}^{[1]}_{2}\left(s^{[1]}_{2}+s^{[2]}_{2}\right)
\!+\!A{h}^{[R,1]}{v}^{[1]}_{3}\left(s^{[1]}_{3}+s^{[2]}_{3}\right)\right\}.
\end{eqnarray}
From the key results using Khintchine-Groshev theorem in
\cite{Motahari1} and \cite{Motahari2}, for any $\epsilon > 0$, a
minimum distance between points in the received signal
constellation point $\mathcal{C}_R$ is given by
\begin{eqnarray}
d_{\textsf{min}}(\mathcal{C}_R)\geq \frac{\kappa
A}{{(2Q)}^{m-1+\epsilon}},
\end{eqnarray}
where $\kappa$ is a constant and $m$ denotes number of rationally
independent integers at the received signal constellation. Thus,
here we can put $m=4$. From \cite{Motahari1} and \cite{Motahari2},
it is not difficult to show that $d_{\textsf{min}}(\mathcal{C}_R)$
increases as $P$ goes to infinity. This means that the error
probability of estimating four desired data symbols
${\hat{s}}^{[1]}_{1}$, ${\hat{s}}^{[2]}_{1}$,
${\hat{s}}^{[1]}_{2}+{\hat{s}}^{[2]}_{2}$, and
${\hat{s}}^{[1]}_{3}+{\hat{s}}^{[2]}_{3}$ becomes zero, and the
relay can decode them with a rate having a multiplexing gain of
$\frac{1}{4}$. Using the estimating data symbols, the relay
creates transmit signal for both destinations. The transmitted
signal at the relay $R$ is
\begin{eqnarray}
{x}^{[R]}&=&B\sum_{j=1}^{4}{v}^{[R]}_{j}{s}^{[R]}_{j},
\end{eqnarray}
where ${s}^{[R]}_{1}={\hat{s}}^{[1]}_{1}$,
${s}^{[R]}_{2}={\hat{s}}^{[2]}_{1}$,
${s}^{[R]}_{3}={\hat{s}}^{[1]}_{2}+{\hat{s}}^{[2]}_{2}$, and
${s}^{[R]}_{4}={\hat{s}}^{[1]}_{3}+{\hat{s}}^{[2]}_{3}$. In
addition, in (26), $B$ represents a power normalizing constant to
satisfy the power constraint at the relay and ${v}^{[R]}_{j}$ is
real coefficient for carrying the integer data symbol
$\hat{s}^{[R]}_{j}$.

The transmit power of the relay $R$ is
\begin{eqnarray}
E\left[{{x}^{[R]}}^2\right]&=&B^2\underbrace{\sum_{j=1}^{4}{{v}^{[R]}_{j}}^2}_{{\eta^{[R]}}^2}\mathbb{E}\left[{{s}^{[R]}_j}^2\right]\leq
B^2{\eta^{[R]}}^2Q^2 \leq P,
\end{eqnarray}
For satisfying the transmit power constraint, we pick $B$ as
\begin{eqnarray}
B = \frac{\bar{\eta} P^{\frac{1}{2}}}{Q},
\end{eqnarray}
where $\bar{\eta}=\frac{1}{\eta^{[R]}}$. The directions of integer
symbols ${s}^{[R]}_{j}$ $j\in\{1,2,3,4\}$ are chosen to cancel the
interference from the unintended source. To accomplish this,
${v}^{[R]}_{j}$ for $j\in\{1,2,3,4\}$ are designed as
\begin{eqnarray}
B{h}^{[2,R]}{v}^{[R]}_{1}{s}^{[R]}_{1}&=&-A{h}^{[2,1]}{v}^{[1]}_{1}{s}^{[1]}_{1}, \quad (\textrm{Destination 2})\nonumber \\
B{h}^{[1,R]}{v}^{[R]}_{2}{s}^{[R]}_{2}&=&-A{h}^{[1,2]}{v}^{[2]}_{1}{s}^{[2]}_{1}, \quad (\textrm{Destination 1})\nonumber \\
B{h}^{[1,R]}{v}^{[R]}_{3}{s}^{[R]}_{3}&=&-A{h}^{[1,2]}{v}^{[2]}_{2}{s}^{[2]}_{2},
\quad (\textrm{Destination 1})\nonumber \\
B{h}^{[2,R]}{v}^{[R]}_{4}{s}^{[R]}_{4}&=&-A{h}^{[2,1]}{v}^{[1]}_{3}{s}^{[1]}_{3},
\quad (\textrm{Destination 2}).
\end{eqnarray}
Since we already picked ${v}^{[1]}_{j}$ and ${v}^{[2]}_{j}$ for
$j\in\{2,3\}$, and for arbitrary given ${v}^{[1]}_{1}$ and
${v}^{[2]}_{1}$ at the sources, we construct ${v}^{[R]}_{j}$ as
\begin{eqnarray}
{v}^{[R]}_{1} &=& \frac{-A{h}^{[2,1]}{v}^{[1]}_{1}}{B{h}^{[2,R]}},
\quad {v}^{[R]}_{2} =
\frac{-A{h}^{[1,2]}{v}^{[2]}_{1}}{B{h}^{[1,R]}}, \nonumber \\
{v}^{[R]}_{3} &=& \frac{-A{h}^{[2,1]}{v}^{[2]}_{2}}{B{h}^{[2,R]}},
\quad {v}^{[R]}_{4} =
\frac{-A{h}^{[2,1]}{v}^{[1]}_{3}}{B{h}^{[1,R]}}.
\end{eqnarray}

\subsubsection{Destinations}
We have shown that two sources ($S_1$ and $S_2$) and the relay $R$
cooperatively design the direction of the integer data symbol so
that each destination $D_i$ can decode its desired data symbols.
Now, we need to check whether each destination $D_i$ can decode
its desired data symbols. The received signal at the destination 1
$D_1$ is given by
\begin{eqnarray}
y^{[1]}&=&{h}^{[1,1]}x^{[1]}+{h}^{[1,2]}x^{[2]}+{h}^{[1,R]}x^{[R]}+n^{[1]} \nonumber \\
&=&{h}^{[1,1]}A\sum_{j=1}^{3}{v}^{[1]}_{j}s^{[1]}_{j}
+{h}^{[1,2]}A\sum_{j=1}^{3}{v}^{[2]}_{j}s^{[2]}_{j}
+{h}^{[1,R]}B\sum_{j=1}^{4}{v}^{[R]}_{j}s^{[R]}_{j}+n^{[1]} \nonumber \\
&\stackrel{(a)}{=}&\left(A{h}^{[1,1]}{v}^{[1]}_{1}+B{h}^{[1,R]}{v}^{[R]}_{1}\right)s^{[1]}_{1}
+\left(A{h}^{[1,1]}{v}^{[1]}_{2}+B{h}^{[1,R]}{v}^{[R]}_{3}\right)s^{[1]}_{2}
\nonumber
\\ &&+\left(A{h}^{[1,1]}{v}^{[1]}_{3}+B{h}^{[1,R]}{v}^{[R]}_{4}\right)s^{[1]}_{3}
+\left(A{h}^{[1,2]}{v}^{[2]}_{3}+B{h}^{[1,R]}{v}^{[R]}_{4}\right)s^{[2]}_{3}+n^{[1]},
\end{eqnarray}
where $(a)$ comes from the fact that
$s^{[i]}_{j}=\hat{s}^{[i]}_{j}$ for $i\in\{1,2\}$ and
$j\in\{1,2,3\}$, and the alignment conditions in (30).

Note that destination 1 $D_1$ wishes to decode
$\tilde{s}^{[1]}_{j}$, $j\in\{1,2,3\}$. From the result in
\cite{Motahari1}, the lower bound of the achievable rate for the
submessage $W^{[1]}_{j}$ can be written as
\begin{eqnarray}
R^{[1]}_{j} &=& I(\tilde{s}^{[1]}_{j};s^{[1]}_{j})  \nonumber \\
&=& H(s^{[1]}_{j})- H(s^{[1]}_{j}|\tilde{s}^{[1]}_{j})  \nonumber \\
&\stackrel{(a)}{\geq}& H(s^{[1]}_{j})- 1 - P^{[1]}_{e,j}\log(|\mathcal{C}_s|)  \nonumber \\
&\stackrel{(b)}{=}& (1-P^{[1]}_{e,j})\log(2Q-1)-1, \quad
j\in\{1,2,3\},
\end{eqnarray}
where $(a)$ follows from Fano's inequality, $(b)$ follows from the
$s^{[1]}_{j}$ is uniformly chosen in the integer set
$[-Q,Q]_{\mathbb{Z}}$, and
$P^{[1]}_{e,j}=\textsf{Pr}(\tilde{s}^{[1]}_{j} \neq
{s}^{[1]}_{j})$. Since the error probability approaches zero,
i.e., $P^{[1]}_{e,j}\rightarrow 0$ as increasing
$P\rightarrow\infty$ as long as we choose the $Q=\gamma
P^{\frac{1-\epsilon}{2(m+\epsilon)}}$. As a result, the achievable
DoF for the submessage $W^{[1]}_{j}$ is
\begin{eqnarray}
d^{[1]}_{j}&=&\lim_{P\rightarrow\infty}\frac{R^{[1]}_{j}(P)}{\frac{1}{2}\log(P)}
\nonumber \\
&=&
\lim_{P\rightarrow\infty}\frac{(1-P^{[1]}_{e,j})\log\left(2\gamma P^{\frac{1-\epsilon}{2(m+\epsilon)}}\right)-1}{\frac{1}{2}\log(P)} \nonumber \\
&=& \frac{1-\epsilon}{m+\epsilon}, \quad j\in\{1,2,3\}.
\end{eqnarray}
Since the received constellation at destination $D_1$, three
integer data symbols are rationally independent, i.e., $m=4$, we
can conclude that
$(d^{[1]}_{1},d^{[1]}_{2},d^{[1]}_{3})=(\frac{1-\epsilon}{4+\epsilon},\frac{1-\epsilon}{4+\epsilon},\frac{1-\epsilon}{4+\epsilon})$
DoF is achieved.

In the same way, the received signal at the destination 2 $D_2$ is
given by
\begin{eqnarray}
y^{[2]}&=&{h}^{[2,1]}x^{[1]}+{h}^{[2,2]}x^{[2]}+{h}^{[2,R]}x^{[R]}+n^{[1]} \nonumber \\
&=&{h}^{[2,1]}A\sum_{j=1}^{3}{v}^{[1]}_{j}s^{[1]}_{j}
+{h}^{[2,2]}A\sum_{j=1}^{3}{v}^{[2]}_{j}s^{[2]}_{j}
+{h}^{[2,R]}B\sum_{j=1}^{4}{v}^{[R]}_{j}s^{[R]}_{j}+n^{[2]} \nonumber \\
&\stackrel{(a)}{=}&\left(A{h}^{[2,2]}{v}^{[2]}_{1}+B{h}^{[2,R]}{v}^{[R]}_{2}\right)s^{[2]}_{1}
+\left(A{h}^{[2,2]}{v}^{[2]}_{2}+B{h}^{[2,R]}{v}^{[R]}_{3}\right)s^{[2]}_{2}
\nonumber
\\ &&+\left(A{h}^{[2,2]}{v}^{[3]}_{3}+B{h}^{[2,R]}{v}^{[R]}_{4}\right)s^{[2]}_{3}
+\left(A{h}^{[2,1]}{v}^{[1]}_{2}+B{h}^{[2,R]}{v}^{[R]}_{3}\right)s^{[1]}_{2}+n^{[2]},
\end{eqnarray}
where $(a)$ comes from the fact that
$s^{[i]}_{j}=\hat{s}^{[i]}_{j}$ for $i\in\{1,2\}$ and
$j\in\{1,2,3\}$, and the alignment conditions in (30).

Using the same argument with Destination 1, we can show that
$(d^{[2]}_{1},d^{[2]}_{2},d^{[2]}_{3})=(\frac{1-\epsilon}{4+\epsilon},\frac{1-\epsilon}{4+\epsilon},\frac{1-\epsilon}{4+\epsilon})$
DoF is achieved. Therefore,
$\sum_{i=1}^{2}\sum_{j=1}^{3}d^{[i]}_{j}=\frac{6}{4}=\frac{3}{2}$
DoF is achievable for the interference channel with the
instantaneous relay when all nodes have a single antenna.

\section{Conclusion} \label{sec:Conclusions}

We explore the DoF of an
interference channel with an instantaneous relay for both single
and multiple antenna systems. By proposing a linear
beamforming scheme and interference-forward relaying scheme, which
is inspired by aligned interference neutralization, we
demonstrate that $\frac{3M}{2}$ DoF is achieved when all nodes
have $M>1$ multiple antennas. The same scheme is applied within the
rational dimensions framework to demonstrate the achievability of
$\frac{3}{2}$ DoF for the single antenna case, i.e., $M=1$. Next we make a few interesting observations based on this DoF result.

\subsection*{\it Observations}
\begin{enumerate}
\item From the point of view of the role of relay, this result is in contrast to
the pessimistic result about conventional relays which do not yield any DoF
gain in the interference channel.
\item The gain in DoF from conventional relays to
instantaneous relays is remarkable because it is quite substantial. The DoF increase by at least 50\% due to instantaneous relaying. This is in contrast to the previous work on instantaneous relaying for single user communication where only modest SNR gains are possible due to instantaneous relaying. This is similar to the recent result in \cite{MaddahTse} which shows that delayed CSIT feedback on a vector broadcast channel can increase the DoF. While the benefits of feedback for broadcast channels have been known for decades, the DoF gain makes it much easier to appreciate the underlying principles.
\item It is also remarkable that such a significant DoF gain is so sensitive to the relative delay between the relayed signal and the direct signal. If the two signals arrive with the same delay, then as we show here, the DoF are improved by at least 50\%, but if the direct signals arrive even one symbol ahead of the relayed signal, then we are in the regime of conventional relays and the DoF gains disappear entirely.

\item Consider the setting where all source and destination nodes have only 1 antenna, but the instantaneous relay node is equipped with 2 antennas. In this case, interestingly, 2 DoF (which is also the outer bound) are easily achieved without any need for alignment at the relay. The relay is able to resolve the source symbols, which allows the relay to neutralize all the interference from both destinations simultaneously, in a manner similar to the cognitive relay channel studied in \cite{Sridharan}.

\item As mentioned earlier, it is remarkable that the DoF gains with instantaneous relaying are possible with only memoryless relaying, i.e., the relay needs no memory of past received symbols.
\end{enumerate}

Finally, an interesting question that remains open is whether $3/2$ is also the DoF outer bound for the interference channel with instantaneous relay, both with and without memory. Other interesting directions would be to study the DoF gain with arbitrary number of antennas at each node and with multiple relay nodes. As explained above, with sufficiently many antennas at the relay nodes, the DoF outer bounds corresponding to full cooperation may be achieved with very simple schemes. This is also in contrast to the pessimistic result of \cite{Cadambe2} for conventional relays, where no DoF gains are possible regardless of the number of antennas at the relay or the number of relay nodes.

\end{document}